\documentclass{IEEEtran}

\usepackage{hyperref}
\usepackage{caption}
\usepackage{indentfirst}

\usepackage{graphicx}
\usepackage{epstopdf}
\graphicspath{{fig/}}
\usepackage{subfigure}
\usepackage{float}
\usepackage{array}
\usepackage{url}
\usepackage{longtable}
\usepackage{rotating}
\usepackage{multirow}
\usepackage{amsmath}
\usepackage{amsfonts}
\usepackage{amssymb}
\usepackage{mathrsfs}
\usepackage{mdwlist}
\usepackage{booktabs}
\usepackage{threeparttable}
\usepackage{makecell}
\usepackage{diagbox}
\usepackage{pifont}

\usepackage{pdflscape}
\usepackage{footnote}

\usepackage[numbers,sort&compress]{natbib}
\usepackage[linesnumbered,boxed,ruled]{algorithm2e}
\usepackage[hang,flushmargin]{footmisc}
\usepackage{lipsum}

\usepackage[T1]{fontenc}
\usepackage[utf8]{inputenc}

\usepackage{color,xcolor}
\definecolor{myred}{RGB}{217,46,127}
\definecolor{mygreen}{RGB}{67,127,127}

\title{AudioSpa: Spatializing Sound Events with Text}

\author{
    {Linfeng Feng, Lei Zhao, Boyu Zhu, Xiao-Lei Zhang, \IEEEmembership{Senior Member, IEEE}, and Xuelong Li, \IEEEmembership{Fellow, IEEE}}

    \thanks{Xiao-Lei Zhang is the corresponding author.}
    \thanks{Linfeng Feng and Lei Zhao contributed equally to this work.}
    \thanks{Linfeng Feng, Lei Zhao, Boyu Zhu and Xiao-Lei Zhang are with the School of Marine Science and Technology, Northwestern Polytechnical University, Xi’an 710072, China, also with the Institute of Artificial Intelligence (TeleAI), China Telecom, Beijing 100033, China, and also with the Research and Development Institute of Northwestern Polytechnical University in Shenzhen, Shenzhen 518063, China (e-mail: fenglinfeng@mail.nwpu.edu.cn, zhao\_lei@mail.nwpu.edu.cn, zhuboyu@mail.nwpu.edu.cn, xiaolei.zhang@nwpu.edu.cn).}
    \thanks{Xuelong Li is with the Institute of Artificial Intelligence (TeleAI), China Telecom, Beijing 100033, China (e-mail: xuelong\_li@ieee.org).}

}
\begin{document}
\maketitle

\begin{abstract}
  Text-to-audio (TTA) systems have recently demonstrated strong performance in synthesizing monaural audio from text. However, the task of generating binaural spatial audio from text, which provides a more immersive auditory experience by incorporating the sense of spatiality, have not been explored yet. In this work, we introduce text-guided binaural audio generation. As an early effort, we focus on the scenario where a monaural reference audio is given additionally. The core problem is to associate specific sound events with their directions, thereby creating binaural spatial audio. The challenge lies in the complexity of textual descriptions and the limited availability of single-source sound event datasets. To address this, we propose AudioSpa, an end-to-end model that applies large language models to process both acoustic and textual information. We employ fusion multi-head attention (FMHA) to integrate text tokens, which enhances the generation capability of the multimodal learning. Additionally, we propose a binaural source localization model to assess the quality of the generated audio. Finally, we design a data augmentation strategy to generate diverse datasets, which enables the model to spatialize sound events across various spatial positions. Experimental results demonstrate that our model is able to put sounds at the specified locations accurately. It achieves competitive performance in both localization accuracy and signal distortion. Our demonstrations are available at \href{https://linfeng-feng.github.io/AudioSpa-demo}{https://linfeng-feng.github.io/AudioSpa-demo}.
\end{abstract}

\begin{IEEEkeywords}
    Text-to-audio, binaural spatial audio, text prompt, binaural source localization.
\end{IEEEkeywords}

\section{Introduction} \label{sec:introduction}
\IEEEPARstart{H}uman binaural hearing enables spatial awareness by processing sound that reaches the ears with time delays, reflections, and blockages influenced by the body. These acoustic cues help the brain interpret spatial information, allowing for sound source localization and interaction with the environment. In real-world environments, multiple sounds often create complex auditory scenes. As a result, generating accurate binaural audio, ranging from simple to complex soundscapes, is essential for providing an immersive auditory experience.

\subsection{Motivation and Challenges}
Real binaural spatial audio is recorded using binaural microphones placed on a dummy head, simulating how human ears receive sound  to recreate the natural listening experience \cite{hammershoi2002methods}. Obviously, capturing such audio requires specialized equipment and skills, making it a technically demanding and costly process. Therefore, it is necessary to study how to generate high-quality binaural audio. Traditional signal processing methods generate binaural audio from monaural audio using a linear time-invariant system, typically modeling interaural time delay and head-related and ear-related filtering \cite{jianjun2015natural}. Alternatively, binaural audio can be generated from monaural audio by convolving with head-related impulse responses (HRIRs). HRIRs can be either recorded or simulated \cite{brinkmann2019cross}, and they serve as the time-domain representation of head-related transfer functions (HRTFs).
% , similar to how room impulse responses (RIR) are used to create reverberant audio \cite{scheibler2018pyroomacoustics}

In recent years, with the development of deep neural networks (DNNs), data-driven audio generation methods have garnered increasing attention. Significant progress has been made in monaural audio generation, with models such as AudioLDM \cite{liu2023audioldm,liu2024audioldm} and Tango \cite{ghosal2023tango,majumder2024tango} leading the way. These models leverage large language models (LLMs) \cite{wu2023large,radford2019language,chung2024scaling,achiam2023gpt} to extract textual information, which is then used to control diffusion models for generating the corresponding audio. For example, AudioLDM employs a typical stage-wise pipeline. It first prepares a pre-trained encoder-decoder structure for compression and reconstruction. During the training stage, the encoder compresses the Mel spectrogram into a latent representation, enabling the diffusion model to generate this compact representation effectively. During the inference stage, the latent representation generated by the diffusion model is passed through the decoder to reconstruct the Mel spectrogram, which is then converted into the final audio using a vocoder. However, current research on spatial audio generation is still limited.

Directly generating binaural audio from a single text modality requires balancing both the generation quality of the sound events and their spatial positioning, making the task highly complex \cite{choi2022proposal}. As an initial work, this paper simplifies the problem by introducing a monaural reference audio, focusing on the relationship between sound events and their spatial locations as described in the text. Related works are the DNN-based monaural-to-binaural audio generation \cite{richard2021neural, leng2022binauralgrad}, which have achieved impressive results in spatial audio synthesis. Similar to traditional methods, these models map an entire monaural audio clip to a specific spatial location by specifying position values such as direction of arrival (DOA) and quaternion. Obviously, the approaches described above are mainly applicable to scenarios where the monaural audio contains a single sound event. However, in real-world situations, audio clips often contain multiple sound sources, such as background noise or various events. Spatializing all sounds to a single position is undesirable. Retaining non-spatialized environmental noises can enhances auditory immersion. For example, the sound of rain suggests the weather, while babble noise conveys the presence of a public space. However, these background noises are typically directionless. Such noise can also convey supplementary information, known as positive-incentive noise \cite{li2022positive}.

Our motivation is to develop a model that spatializes specific sound events, while preserving the undirected nature of the remaining sounds in the monaural reference audio. In this paper, we use text to describe the spatial locations of the sound events. To achieve this, several challenges must be addressed. First, how to effectively integrate the text tokens with the acoustic features for multimodal fusion. Second, the scarcity of suitable datasets presents a significant obstacle. Existing large-scale sound event datasets, such as AudioSet \cite{gemmeke2017audio}, are multi-source. Therefore, we need to design a method to construct a large number of monaural-binaural-text pairs using relatively small amounts of single-source data. Additionally, it is essential to design evaluation metrics specifically for assessing spatial perception.

\subsection{Goals and Contributions}
Based on the analysis above, we propose a monaural-to-binaural generation model, along with a binaural localization model and a data augmentation strategy. We demonstrate the effectiveness of our proposed method across experiments in diverse scenarios. The contributions can be summarized as follows:
\begin{itemize}
    \item
    \textbf{We propose AudioSpa, an end-to-end monaural-to-binaural generation model.} It is a time-domain waveform mapping model, with the backbone consisting of 1D convolutional residual blocks. The model takes embeddings extracted by a LLM as conditions, directly mapping the monaural waveform to the text-specified binaural waveform.

    \item
    \textbf{We propose a binaural localization model to evaluate generative performance.} The generated binaural audio is processed through the localization model, and the resulting DOA is compared with the DOA described in the text. The angle error is recorded as one of the evaluation metrics.

    \item
    \textbf{We propose a data augmentation strategy to address the data scarcity issue.} During training, this strategy dynamically mixes sound events, spatial information, and noise, creating diverse sound scenes. This allows the model to effectively learn the relationships between these varied sounds and the text descriptions, increasing the generalization ability.
\end{itemize}

This paper is organized as follows. Section~\ref{sec:re} presents related work. Section~\ref{sec:spa} describes the proposed AudioSpa in detail. Section~\ref{sec:local} presents the design of the localization model. Section~\ref{sec:data} describes the datasets and the data augmentation strategy. Section~\ref{sec:exp} outlines the experimental setup and results. Section~\ref{sec:dis} discusses the limitation and future work. Finally, Section~\ref{sec:con} concludes the current work.

\section{Related Work}   \label{sec:re}
\subsection{Sound Generation}
First, we discuss monaural sound generation. Recent advancements in universal sound generation, such as AudioLDM \cite{liu2023audioldm,liu2024audioldm} and Tango \cite{ghosal2023tango,majumder2024tango}, have already been discussed. In addition, related monaural sound generation works can be categorized based on input types (text, image, video) and output types (speech, music, universal sound). These input-output combinations lead to various tasks, such as text to speech \cite{tan2024naturalspeech}, text to music \cite{chen2024musicldm}, image to music \cite{wang2023continuous}, video to audio \cite{ghose2023foleygan}, and text to audible video \cite{liu2023sounding}, etc.

Regarding binaural sound generation, both MusicGen \cite{copet2024simple} and Stable Audio \cite{evans2024stable} can produce binaural audio without relying on monaural reference audio. However, they lack the ability to specify the precise locations of sound events, i.e., the spatial perception is randomly generated. In contrast, with monaural reference audio, the spatial control information can be classified into different input modalities. The input in \cite{richard2021neural, leng2022binauralgrad} is location values (DOA or quaternion), while the input in \cite{gao20192, li2024cross, li2024cyclic} is video, aligning the spatial perception of sound events with the visual scene.

\subsection{Sound Separation and Localization}
The monaural-to-binaural approach in \cite{gao20192, li2024cross, li2024cyclic} follows a paradigm where monaural audio is treated as a mixture of binaural audio. It separates the two channels from monaural audio using a method similar to speech separation \cite{wang2018supervised}. Their training objective is a complex mask or complex ideal ratio mask (cIRM) \cite{williamson2016complex} derived from speech separation. From the input-output perspective, the text-driven monaural-to-binaural generation is similar to text-driven target sound separation \cite{liu2022separate,liu2024separate,ma2024clapsep,ma2024language}, both being audio-to-audio conversion tasks controlled by text.

From the time delay perspective, the input audio used in \cite{richard2021neural, leng2022binauralgrad} involves the warping of the monaural audio, which is a technique of adding time delay from the sound source to the ears. This is similar to the concept of direct sound in dereverberation tasks \cite{zhao2024multi, guo2024graph}. From a spatial information perspective, the monaural-to-binaural task can be seen as generating spatial information for sound, while sound source localization \cite{feng2023soft,feng2024learning,feng2025eliminating} is the task of recognizing spatial information from sound.
% Therefore, monaural-to-binaural generation can be viewed as a reverse dereverberation process, where the warping is enhanced with attenuation, reflections, and scattering effects.

\section{AudioSpa}   \label{sec:spa}

\begin{figure*}[t]
    \centering
    \includegraphics[width=1.0\textwidth]{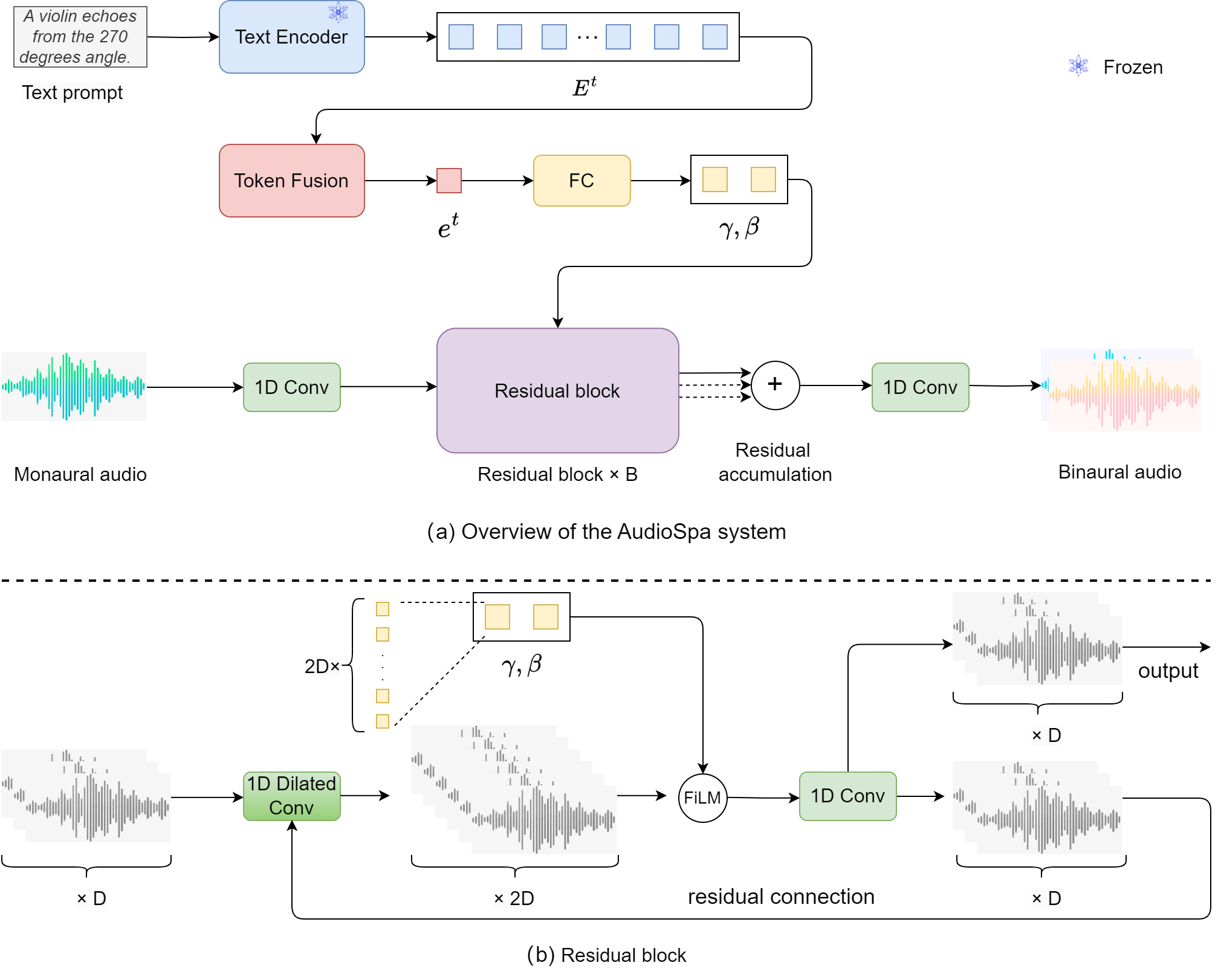}
    \caption{The model architecture of AudioSpa, which takes input text and monaural audio and outputs binaural audio in an end-to-end manner. For simplicity, we omit the activation functions.}
    \label{fig:mono2bi}
\end{figure*}

In this section, we outline the pipeline of AudioSpa. First, we discuss the construction of input-output data pairs. Next, we describe the backbone architecture used for audio modeling. This is followed by the detail of the text encoder and the multimodal fusion approach. Finally, we describe the loss function.

\subsection{Signal Model} \label{subsec:signal}
Firstly, we consider the scenario of a single sound source in a clean environment. Given a raw monaural audio $\mathbf{x}^\mathrm{raw} \in \mathbb{R}^{1 \times N}$ and a recorded HRIR $\mathbf{r} \in \mathbb{R}^{2 \times N^r}$, the relationship between the mixture signal $\mathbf{y} \in \mathbb{R}^{2 \times N}$, the direct-path signal $\mathbf{s} \in \mathbb{R}^{2 \times N}$, and the reverberation $\mathbf{h} \in \mathbb{R}^{2 \times N}$ from the listener’s body are related as follows:
\begin{equation}\label{eq:signal_bi}
    \begin{aligned}
        \mathbf{y} &= \mathbf{x}^\mathrm{raw} * \mathbf{r} \\
         &=  \mathbf{s} + \mathbf{h}
    \end{aligned}
\end{equation}
where $*$ denotes the linear convolution operator, and after convolution, only the first $N$ samples of $\mathbf{y}$ are retained. Thus, we have obtained the training objective $\mathbf{y}$. In addition, the spatial information in the HRIR will be used to generate text prompts.

Next, to construct the data pair, the input signal $\mathbf{x}$ is derived from the raw audio $\mathbf{x}^\mathrm{raw}$ by applying a time delay $\tau$. The delay $\tau$ is calculated based on the distance $d$ between the sound source and the listener’s head, the speed of sound $c$, and the sampling rate $f_s$, using the following formula:
\begin{equation}\label{eq:tau}
\tau = \frac{d}{c} \cdot f_s
\end{equation}
where $\tau$ is rounded to the nearest integer. The input signal $\mathbf{x}$ is then obtained as follows:
\begin{equation}\label{eq:x_tau}
    \mathbf{x}(n) =
    \left\{\begin{array}{ll}
      \mathbf{x}^\mathrm{raw}(n-\tau),& \mbox{ if   } n > \tau \\
      0,& \mbox{ otherwise}
    \end{array}\right.,\quad \forall n = 1,\ldots, N
\end{equation}
Thus far, the input-output data pair $\mathbf{x}$ and $\mathbf{y}$ has been constructed. The delay of $\mathbf{x}$ approximates the average of the two channels of the direct-path signal $\mathbf{s}$.
% However, $\mathbf{s}$ is affected by head and ear occlusion, which may lead to volume differences between the two channels. Therefore, no amplitude manipulation is applied to $\mathbf{x}$.

Secondly, considering the scenario of a single sound source in the environment with ambient noise, we set that the noise $\mathbf{v} \in \mathbb{R}^{1 \times N}$ is unchanged in both the input and output of the model, with the noise received by both ears being identical. Thus, the data pair becomes $\mathbf{x} + \mathbf{v}$ and $\mathbf{y} + \mathbf{v}$.

% For simplicity, we define the time delay of $\mathbf{x}$ as the time required for the sound to travel from the source to the center of the ears.
% . Due to the head and ear occlusion effects, there may be significant volume differences between the two channels of $\mathbf{s}$. Therefore, when constructing the data, we do not adjust the volume of $\mathbf{x}$, but only align the time delay

\subsection{Audio Backbone} \label{subsec:backbone}
As illustrated in Fig.~\ref{fig:mono2bi}, the backbone component of the entire model is a time-domain architecture. This structure transforms time-domain monaural waveforms into binaural waveforms through the use of multiple stacked 1D convolutional layers. It has been proven that time-domain models can perform powerfully in both sound separation and synthesis \cite{luo2019conv,kong2021diffwave}.

For clarity, We first discuss the overview of the backbone. The architecture starts with a 1D convolutional layer serving as the input projection, which maps the input $x$ into a higher-dimensional space, yielding $\mathbf{E}^{x} \in \mathbb{R}^{D \times N}$. Subsequently, $\mathbf{E}^{x}$ is processed through $B$ residual blocks, where it undergoes intricate transformations that ultimately result in $\mathbf{E}^{y} \in \mathbb{R}^{D \times N}$. Finally, the output projection transforms $\mathbf{E}^{y}$ into the desired output $\hat{\mathbf{y}} \in \mathbb{R}^{2 \times N}$. The residual channel $D$ and the number of residual blocks $B$ are the two tunable hyperparameters in the backbone.

Returning to the residual block, to increase the receptive field for modeling long audio sequences, we use the the dilated convolution layers. The dilation (i.e., spacing between kernel elements) of the dilated convolution changes with the depth of the block. In this work, the dilation values cycle through $\{1, 2, \dots, 512\}$. Additionally, the output channels of the dilated convolution are set to twice the input channels. Specifically, the input $\mathbf{E}^{x} \in \mathbb{R}^{D \times N}$ produces an output $\mathbf{E}^a \in \mathbb{R}^{2D \times N}$. This doubling is intended to split the output into two parts, which are then passed through sigmoid and tanh activations, respectively. Afterward, the output undergoes multimodal fusion using a feature-wise linearly modulated (FiLM) layer \cite{perez2018film}. Finally, a 1D convolution layer adjusts the number of channels, with half of the output channels serving as the output of the current residual block and the other half as the input to the next residual block, as shown in Fig.~\ref{fig:mono2bi}.

\subsection{Text Encoder and Muitimodal Fusion} \label{subsec:text}
In this paper, we use FLAN-T5 \cite{chung2024scaling} as the text encoder $f_{text}(\cdot)$ for extracting textual features. This text encoder is a pre-trained LLM and remains frozen throughout training. A text prompt describing the spatial location of the sound events is processed by $f_{text}(\cdot)$, producing textual features $\mathbf{E}^{t} \in \mathbb{R}^{N^t \times D^t}$, where $N^t$ represents the number of tokens, and $D^t$ is the embedding dimension of each token.

To achieve multimodal fusion of text and audio, we employ a FiLM layer \cite{perez2018film} after the dilated convolution layer in each residual block. First, we need compressing $\mathbf{E}^{t} \in \mathbb{R}^{N^t \times D^t}$ into a single token $\mathbf{e}^{t} \in \mathbb{R}^{D^t}$. One possible approach is to directly select the first token from $\mathbf{E}^{t}$ \cite{liu2024separate}, based on the reasoning that the LLM inherently captures contextual relationships among tokens, enabling the first token to carry contextual information. Nevertheless, we opt to let the model autonomously derive $\mathbf{e}^{t}$ by employing the following fusion multi-head attention (FMHA). The subsequent experimental results will prove that the generation quality using FMHA is much better than directly selecting the first token. Here are the details of FMHA:
\begin{equation}\label{eq:mha}
  \mathbf{e}^{t}=\mathrm{FMHA}(\mathbf{E}^t)=\operatorname{Concat}\left(\mathbf{O}_{1}, \ldots, \mathbf{O}_{h}, \ldots, \mathbf{O}_{H}\right) \mathbf{W}^{o}
\end{equation}
where $H$ denotes the number of attention heads, $\mathbf{W}^{o} \in \mathbb{R}^{D^t \times D^t}$ is the learnable output projection matrix, and $\mathbf{O}_h$ with $h = 1, \dots, H$ represents the output of each head:
\begin{equation}\label{eq:fmha}
    \mathbf{O}_h=\operatorname{Softmax}\left(\frac{\mathbf{q}^T\left(\mathbf{W}_h^k \mathbf{E}^t\right)}{\sqrt{D^t / H}}\right)\left(\mathbf{W}_h^v \mathbf{E}^t\right), \quad \forall h=1, \ldots H
\end{equation}
where $\mathbf{q} \in \mathbb{R}^{\frac{D^t}{H} \times 1}$, $\mathbf{W}_h^k \in \mathbb{R}^{\frac{D^t}{H} \times N^t}$ and $\mathbf{W}_h^v \in \mathbb{R}^{\frac{D^t}{H} \times N^t}$ are learnable projection matrices.

Then, $\gamma, \beta \in \mathbb{R}^{2D}$ are the modulation parameters obtained from:
\begin{equation}\label{eq:film2}
 (\gamma, \beta) = f_{film}(\mathbf{e}^{t})
\end{equation}
where $f_{film}(\cdot)$ is a neural network, which in this work is modeled using two fully connected layers followed by a PReLU activation \cite{he2015delving}.

The modulation parameters $\gamma, \beta \in \mathbb{R}^{2D}$ are applied to the acoustic features $\mathbf{E}^a \in \mathbb{R}^{2D \times N}$ using the FiLM layer as follows:
\begin{equation}\label{eq:film}
    \mathrm{FiLM}(\mathbf{E}^a_n|\gamma, \beta) = \gamma \odot \mathbf{E}^a_n + \beta, \quad \forall n=1, \ldots N
\end{equation}
where $\odot$ is the Hadamard product operator.

\subsection{Loss Function} \label{subsec:loss}
In this paper, the designed AudioSpa is a regression model. Following the state-of-the-art sound separation techniques \cite{luo2023music, wang2023tf}, we use the L1 loss function. Specifically, let the ground-truth be $\mathbf{y}$ and the model prediction be $\hat{\mathbf{y}}$, then the loss function is:
\begin{equation}\label{eq:loss}
    \mathcal{L}^{\mathrm{L1}} = |\hat{\mathbf{y}} - \mathbf{y}|_{1}
\end{equation}

\section{Binaural Localization}   \label{sec:local}
\begin{figure*}[t]
    \centering
    \includegraphics[width=0.72\textwidth]{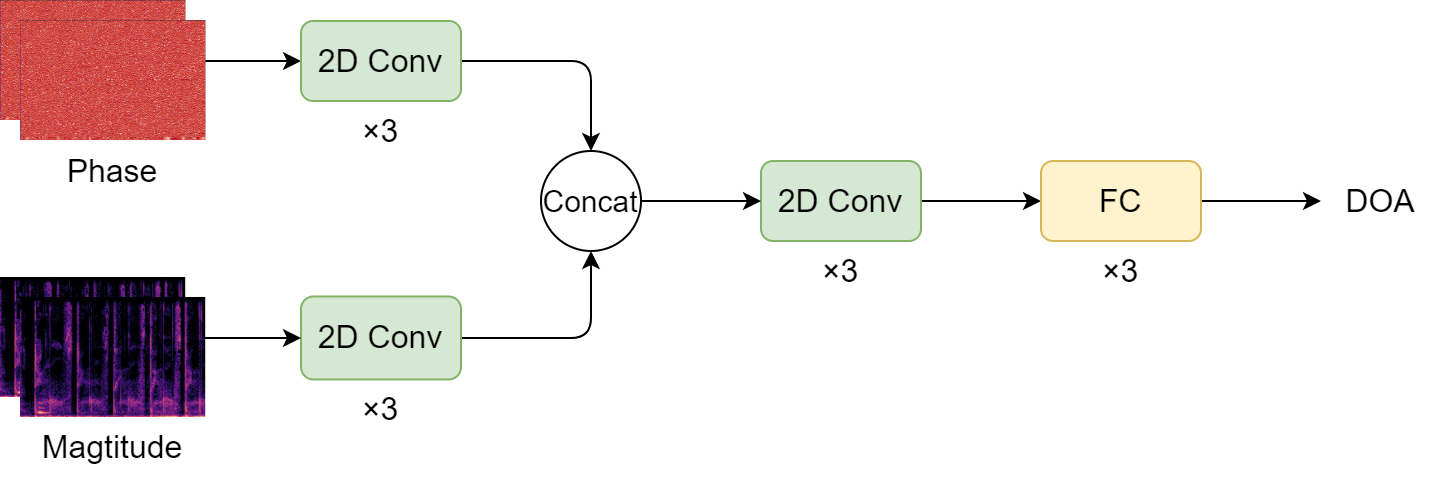}
    \caption{The architecture of the binaural localization model. For simplicity, we omit the activation functions.}
    \label{fig:local}
\end{figure*}

To evaluate the spatial quality of the generated binaural audio, we design a DNN-based binaural localization model. Firstly, the model achieves a localization error close to zero when the input is the ground-truth binaural audio. Consequently, when the input is the generated audio, the resulting localization error can be approximately interpreted as the angular error of the generated audio itself, reflecting the spatial quality.

\subsection{Model Design}
First, we begins with the input features. The human ears can be viewed as a two-channel fixed microphone array, where phase spectrogram is commonly used as a feature map for sound source localization in fixed arrays \cite{feng2025eliminating}. Unlike two-channel linear arrays, the sound magnitudes received by the left and right ears may differ significantly due to head shadowing. As a result, binaural localization often relies on features like interaural phase difference (IPD) and interaural level difference (ILD) \cite{grumiaux2022survey}.  The computation of these interaural difference features is linear. For example, ILD can be calculated as the level of the left ear minus that of the right ear. In contrast, DNNs are designed to stack linear and nonlinear operations, which can learn more flexible feature representations. Therefore, to achieve this, we directly input the raw phase and magnitude spectrograms of the binaural signals into the convolutional layers. In simple terms, we set the input channel size of the first convolutional layers to 2, corresponding to the two ears.

Based on the above analysis, our binaural localization model is inspired by the phase model in \cite{feng2025eliminating} and follows the classification-based design. The key differences are as follows: (i) The original model processes only the phase spectrograms, whereas our current model introduces an additional input branch to handle the magnitude spectrograms. (ii) In the current model, the convolutional channels operate directly on the microphone channels (i.e., the binaural inputs).

As shown in Fig.~\ref{fig:local}, the model applies three consecutive convolutional layers to the phase and magnitude spectrograms respectively, producing two feature maps, $\mathbf{E}^{pha'} \in \mathbb{R}^{c' \times f' \times t'}$ and $\mathbf{E}^{mag'} \in \mathbb{R}^{c' \times f' \times t'}$. These features are concatenated to form $\mathbf{E}^{mix'} \in \mathbb{R}^{2c' \times f' \times t'}$, which is further fused through another three convolutional layers, resulting in $\mathbf{E}^{mix''} \in \mathbb{R}^{c'' \times f'' \times t''}$. The resulting tensor is flattened into a vector and passed through three fully connected layers to produce $\hat{\mathbf{p}} \in [0, 1]^{I}$, representing the probability distribution of the sound sources across $I$ possible locations. In this work, we divide the azimuth plane into $I=36$ classes, corresponding to angles of $\{0, 10, \dots, 350\}$.

\subsection{Loss Function}
To enable this model to support both single and multiple sound source localization, we adopt a multi-label classification strategy. In this approach, the probability of a location having a sound source is set to 1 if a source is present, and 0 otherwise. Therefore, we use binary cross-entropy (BCE) loss function. Specifically, let the ground truth be $\mathbf{p}$ and the model's prediction be $\hat{\mathbf{p}}$, the loss function is defined as:
\begin{equation}\label{eq:bce}
    \mathcal{L}^{\mathrm{BCE}}=-\sum_{i=1}^I p_i \log \hat{p}_i + (1-p_i) \log (1-\hat{p}_i)
\end{equation}

\section{Datasets and Augmentation}   \label{sec:data}
In this section, we provide a detailed description of the datasets used for AudioSpa. Due to the limited availability of single-source sound event datasets, we adopt an on-the-fly data construction approach in the training stage. This involves mixing the real-world HRIRs, the sound event datasets, and text prompts. Additionally, since some of the experimental scenarios in this paper involve environmental noise, this section also includes a description of the noise datasets used. All datasets used were resampled to 24kHz.

\subsection{Datasets}

\subsubsection{HRIRs}
% Since this study does not involve personalized customization,
We selected the HUTUBS HRTF database \cite{brinkmann2019hutubs}. For simplicity, we use the measured HRIRs from a single subject. We chose the ``pp96'' subject, whose head dimensions are 14.81 cm in width, 21.55 cm in height, and 18.73 cm in depth. This set of recorded HRIRs offers 440 spatial points on the 3D spherical surface. Since this work only considers the azimuth angles and excludes the elevation angles, we selected the 36 spatial points with an elevation of 0° for constructing binaural audio used for training and evaluation.

\subsubsection{Sound Event Datasets}
It is important to clarify that, in this paper, we focus on scenarios where only one sound source is present at a single location. Therefore, we exclude some large datasets like AudioSet \cite{gemmeke2017audio}, as many of their audio samples already contain a mixture of multiple sound events. Instead, we selected the UrbanSound8K \cite{salamon2014dataset}, ESC-50 \cite{piczak2015esc}, FSD50K \cite{fonseca2021fsd50k}, and MUSDB18-HQ \cite{rafii2019musdb18} datasets, as their audio consists of single sound events.

UrbanSound8K, ESC-50, and FSD50K are similar in that they contain a wide variety of sound types. UrbanSound8K contains 8,732 audio clips across 10 categories, covering common sounds in real-world urban environments, such as car horns, alarms, dog barks, and more. ESC-50 consists of 2,000 audio clips across 50 categories, with 40 samples per category. The dataset features sounds from various environments, including both natural and man-made sounds, such as animal noises, weather sounds, and tools. FSD50K includes 51,197 audio clips across 200 categories, representing a wide range of everyday environmental sounds, including cars, birds, instruments, and natural sounds. However, some of the audio clips contain long periods of silence. To address this, we applied voice activity detection (VAD) \cite{hwang2023torchaudio} to trim each audio file. After trimming the silence at the beginning and end, audio clips shorter than 0.5s or longer than 30s were discarded. After processing, the audio from these three datasets is mixed and randomly split into training, validation, and test sets with a ratio of 90\%, 5\%, and 5\%, respectively.

MUSDB18-HQ is a dataset for music source separation \cite{luo2023music}, consisting of 150 songs, each with 4 independent tracks: bass, drum, vocal, and other. It contains two subsets: a training set with 100 songs and a test set with 50 songs. In this paper, we split the original test set into two equal parts for validation and testing. Since each song is several minutes long, we first trim long silent segments from the individual tracks, then divide them into 6-second segments. After processing, there are approximately 20,000 segments. Combined with the previously mentioned sound event datasets, the data for training, validation, and testing consists of roughly 65,000, 3,500, and 3,500 segments, respectively.

\subsubsection{Noise Datasets}
The environmental noise datasets used in this paper are sourced from the DNS5 Challenge \cite{dubey2023icassp} and WHAM! \cite{wichern2019wham}. The DNS5 Challenge \cite{dubey2023icassp} aims to joint denoising, dereverberation, and suppression of interfering talkers. Its noise dataset contains approximately 180 hours of audio, sourced from AudioSet \cite{gemmeke2017audio} and FreeSound \cite{fonseca2017freesound}, with sound conditions derived from crowd-sourced or YouTube sources. WHAM! \cite{wichern2019wham} is a noisy speech separation dataset. We use only the real ambient noise samples from this dataset, totaling around 70 hours, with recordings from 4 urban environments. In total, the noise datasets used amount to approximately 250 hours.

\subsection{Data Augmentation}
The proposed on-the-fly data augmentation algorithm utilizes a set of HRIRs, a mixed sound event dataset, a mixed noise dataset, and a collection of descriptive templates. The descriptive templates consist of a set of predefined sentences, such as ``At [azimuth] degrees, the [sound event] rings out,'' where placeholders such as ``sound event'' and ``azimuth'' are to be filled with specific values. In this paper, we use GPT-4o \cite{achiam2023gpt} to generate the descriptive templates.
% , as additional similar descriptions would not introduce significant variation.

During training, each epoch corresponds to one iteration over the sound event dataset, where each sound event audio is used once. After selecting a sound event audio, the algorithm randomly chooses an HRIR and a descriptive template. The chosen template is then filled with relevant details, such as the azimuth angle and sound event, to generate text-monaural-binaural data pairs. Specifically, many samples in the previously mentioned sound event datasets have multiple event labels. For example, an audio clip of a bass performance might be labeled both as ``bass'' and ``music.'' For such multi-label samples, we randomly select one label during training to construct the text prompt.
% The binaural data is synthesized using the HRIR, while the monaural data is directly derived from the selected sound event audio.

% we ensure that the average energy of the two binaural channels equals the energy of the monaural audio.
Next, we discuss scenarios with noise. Based on the aforementioned text-monaural-binaural data pairs,  a noise segment of the same length as the monaural audio is randomly selected from the noise dataset. This noise is then added to both the monaural and binaural audio tracks at the same signal-to-noise ratio (SNR). The SNR for the binaural audio is calculated based on its average energy.

This augmentation method introduces variation in both the textual descriptions, sound events, spatial, and environmental aspects, thereby enhancing the model's generalization ability by generating a diverse range of training examples on-the-fly.

\section{Experiments}   \label{sec:exp}
\subsection{Training Details}
By default, the residual channel $D$ in the audio backbone of Section~\ref{subsec:backbone} is set to 64, and the number of residual blocks $B$ is set to 30. The dilation for the dilated convolutions is configured to cycle every 10 residual blocks, with the dilation values cycling through $\{2^0, 2^1, \dots, 2^9\} = \{1, 2, \dots, 512\}$. Under the default setting, this results in 3 complete cycles, ensuring that the model can handle long audio sequences. In this work, the chosen text encoder is FLAN-T5 \cite{chung2024scaling}. By default, we use FLAN-T5-large, which has a model size of 783M parameters, and the output dimension of a token, $D^t$, is 1024. Additionally, in ablation experiments, we use the smaller FLAN-T5-base, which has a model size of 248M parameters and a corresponding $D^t$ of 768. However, regardless of the model, directly feeding the tokens into the FMHA for token fusion results in excessive computational cost, as the attention module's complexity scales quadratically. Therefore, we introduce two fully connected layers to compress the token dimension to $2D$, and then perform token fusion. With the default settings, $2D = 128$.

In each training epoch, we sample a 4-second segment from each monaural audio to generate data pair. Before using them for training, we normalize the sample variance of each monaural segment to 1.0 and apply the same scaling factor to the target binaural audio. This normalization method has been shown to effectively improve model performance in state-of-the-art speech separation \cite{wang2023tf}. The binaural sound source localization model uses the same data as the generative model, with the training, validation, and test set splits being exactly the same.

For all experiments, we used the AdamW optimizer with a maximum of 50 training epochs. The learning rate was initialized at $10^{-3}$. At the end of each epoch, we evaluated the validation loss and based on that, we set up a learning rate decay strategy and early stopping. The learning rate decay strategy has a patience of 3 epochs for the validation loss; if no improvement is observed within 3 epochs, the learning rate is reduced by half, with a minimum value of $10^{-4}$. Early stopping has a patience of 10 epochs for the validation loss; training is stopped if no improvement is seen beyond 10 epochs. Finally, the checkpoint with the lowest validation loss is selected as the best model for testing.
% \cite{loshchilov2019decoupled}

\subsection{Comparison of Methods}
For comparison with previous methods, we can only perform comparisons in single-source scenarios with a clean environment. In more complex environments, we only conduct ablation experiments to compare the effects of model size, the use of LLMs, token fusion methods, and whether data augmentation. Specifically, we will compare the ground-truth binaural audio to assess the performance of the localization model. Below are the previous methods compared in a clean environment:
% We will also compare by replicating the monaural audio into a two-channel format.
\begin{itemize}
    \item \textbf{Digital signal processing (DSP)} \cite{brinkmann2019hutubs}: Using traditional DSP, HRIRs are simulated based on the head, ear, and shoulder parameters of the recording subject, and then convolved to generate binaural audio.
    \item \textbf{WarpNet} \cite{richard2021neural}: It stacks a neural time warping module with a temporal convolutional network to generate binaural audio.
    \item \textbf{BinauralGrad} \cite{leng2022binauralgrad}: It uses position information and monaural audio as conditional embeddings to control a diffusion model that generates the corresponding binaural audio. Specifically, it proposes a two-stage generation scheme, so both its one- and two-stage methods are used as baselines.
\end{itemize}

\subsection{Evaluation Metrics}
For spatial evaluation, we adopt common metrics from sound source localization \cite{feng2025eliminating}: DOA mean absolute error (MAE) and classification accuracy (ACC).
The ACC is self-explanatory. For the DOA MAE, however, special attention is needed due to the cyclic nature of angles on the horizontal plane. Given a sample with the ground-truth DOA \(\theta\) and the predicted DOA \(\hat{\theta}\), the MAE is computed as follows:
\begin{equation}\label{eq:mae}
    \mathrm{MAE}(\circ)= \mathrm{min}(|\hat \theta - \theta|_1, 360-|\hat \theta - \theta|_1)
\end{equation}

Additionally, since this work uses ground-truth binaural audio as a reference, we evaluate the generation quality using signal-to-distortion ratio (SDR) and scale-invariant signal-to-distortion ratio (SISDR). Higher values for these metrics indicate that the model generates binaural audio closer to desired outcome. Specifically, SISDR is designed as an SDR that disregards volume effects. Since interaural level difference influences perceived spatial attributes, a detail should be considered when calculating the scaling factor for SISDR: we treat the entire binaural audio as a whole. Therefore, we flattened the binaural audio \(\mathbf{y}\) and \(\hat{\mathbf{y}}\) into vectors of length \(2N\). The SDR is computed as follows:
\begin{equation}\label{eq:sdr}
    \mathrm{SDR}(\mathrm{dB})= \frac{1}{2N} \sum_{n=1}^{2N}10\mathrm{log}_{10} \frac{|y_n|_2^2}{|y_n - \hat{y}_n|_2^2}
\end{equation}
and the SISDR is computed as follows:
\begin{equation}
    \left\{\begin{array}{l}
    \bar{\mathbf{y}} = \frac{\hat{\mathbf{y}}^T \mathbf{y}}{|\mathbf{y}|_2^2} \\
    \mathrm{SISDR}(\mathrm{dB})= \frac{1}{2N} \sum_{n=1}^{2N}10\mathrm{log}_{10} \frac{|\bar{y}_n|_2^2}{|\bar{y}_n - \hat{y}_n|_2^2}
    \end{array}\right.
\end{equation}
where \( |\cdot|_2^2 \) represents the squared L2 norm. As seen from the definition of SISDR, both channels of the binaural audio are scaled synchronously. The physical interpretation of this is that changing the overall volume of the binaural audio does not alter its spatial perception.

\subsection{Main Results}
\subsubsection{Single Source in Clean Environment}
\begin{table}[t]
    \centering
    \caption{Results on the single-source data, where the environment is clean.}
    \scalebox{0.96}{\begin{tabular}{ccccc}
      \toprule
      Method & MAE ↓ & ACC (\%) & SDR (dB) & SISDR (dB) \\
      \midrule
      Ground-truth & 0.02  & 99.97 & $\infty$   & $\infty$ \\
      Mono  & 88.98 & 3.06  & -4.07 & -19.98 \\
      \midrule
      DSP   & 22.18 & 70.25 & -0.76 & -13.89 \\
      WarpNet \cite{richard2021neural} & 0.20  & 99.67 & 29.68 & 31.42 \\
      BinauralGrad \cite{leng2022binauralgrad} (1) & 0.12  & 99.49 & 30.62 & 32.04 \\
      BinauralGrad \cite{leng2022binauralgrad} (2) & 0.11  & 99.53 & 31.13 & 32.19 \\
      \textbf{AudioSpa} & 0.03  & 99.83 & 41.61 & 42.13 \\
      \bottomrule
      \end{tabular}}
    \label{tab:1-clean}%
  \end{table}%

The main objective of the experiments is to validate the model's ability to generate spatial audio in a simple scenario. As shown in Table~\ref{tab:1-clean}, using ground-truth audio as input, we can observe that the source localization model performs accurately in the clean environment, with a DOA MAE of only 0.02 degrees and a classification accuracy of 99.97\%. When comparing with monaural audio, the DOA MAE increases to 88.98 degrees, and the classification accuracy drops to 3.06\%. This essentially means a complete lack of spatial perception, as the expected angle difference between two random points in space is 90 degrees.

It is worth noting that while DSP methods show a gap in evaluation metrics compared to DNN methods, they still deliver a reasonably good auditory experience, with spatial perception closely resembling that of ground-truth audio \footnote{\href{https://linfeng-feng.github.io/AudioSpa-demo}{https://linfeng-feng.github.io/AudioSpa-demo}}. In the clean environment, the DNN methods show minimal differences, with SDR above 30 dB, making them indistinguishable to human ears. AudioSpa achieves a DOA MAE of 0.03 degrees and a DOA accuracy of 99.83\%, demonstrating its effectiveness in assigning spatial attributes to sound events in such a simple scenario.

\subsubsection{Single Source in Noisy Environment}

\begin{table}[t]
    \centering
    \caption{Results on the single-source data under noisy environments with varying SNR (in dB).}
    \scalebox{1.0}{\begin{tabular}{ccccc}
        \toprule
        SNR of Input & MAE ↓ & ACC (\%) & SDR (dB) & SISDR (dB) \\
        \midrule
        Ground-truth & 0.24  & 99.52 & $\infty$ & $\infty$ \\
        Mono  & 89.60 & 2.60  & -1.56 & -12.12 \\
        \midrule
        0 dB     & 8.00  & 89.82 & 4.15  & 0.92 \\
        5 dB    & 7.21  & 90.51 & 6.91  & 5.54 \\
        10 dB    & 6.71  & 91.37 & 9.44  & 8.79 \\
        15 dB   & 5.64  & 92.87 & 11.32 & 11.01 \\
        $\infty$ dB & 3.47  & 95.64 & 14.21 & 14.01 \\
        \bottomrule
        \end{tabular}}%
    \label{tab:1-noisy}%
  \end{table}%

The main objective of the experiments is to validate the capability of the proposed AudioSpa model to spatialize a specified sound event while preserving the undirected nature of background sounds. The tested model was trained on data with SNR values uniformly distributed in the range of $[0, 15]$ dB. In Table~\ref{tab:1-noisy}, both the ground-truth binaural audio and monaural audio also have SNR values uniformly distributed within $[0, 15]$ dB. It can be observed that our binaural localization model achieves high accuracy even in noisy environments, with a DOA MAE of only 0.24 degrees and an ACC of 99.52\%. Notably, the ``signal'' used for calculating SDR and SISDR refers to the noisy binaural audio. As a result, the SDR and SISDR values for the ground truth compared to itself are $\infty$ dB. For this reason, compared to Table~\ref{tab:1-clean}, the SDR of monaural audio increases. This is because the only difference between the monaural audio and the binaural audio is the target sound event, while the remaining sounds in both audio tracks are identical.

As the SNR of the target signal increases (perceptually resembling the target source moving closer to the listener), the model's performance improves in both DOA and SDR metrics. Notably, even in scenarios where the target signal's SNR is as low as 0 dB (i.e., the target signal's energy is equal to that of the background noise), the localization performance remains acceptable, with a DOA MAE of just 8 degrees and a classification accuracy of 89.82\%. However, the SDR is relatively low, indicating room for improvement in perceptual quality. When the SNR reaches 10 dB, the overall performance becomes satisfactory, with improved localization accuracy and perceptual quality, closely resembling the ground-truth binaural audio. Since the training data always includes noise, AudioSpa's performance on clean data is slightly inferior to its performance on matched data in Table~\ref{tab:1-clean}. Nevertheless, an SDR of 14.21 dB still delivers a perceptually good listening experience.

\subsection{Ablation Study}
This subsection focuses on ablation studies to examine the effects of the text encoder, model parameter size, and data augmentation on model performance. To ensure sufficient differentiation in the experiments, the test environment is noisy. All experiments are conducted on a test dataset with SNR values uniformly distributed in the range of $[0, 15]$ dB.

\begin{table}[t]
    \centering
    \caption{The impact of different text encoders and token fusion methods on the performance of AudioSpa.}
    \scalebox{0.90}{\begin{tabular}{cccccc}
      \toprule
      LLM   & Token Fusion & MAE↓  & ACC (\%) & SDR (dB) & SISDR (dB) \\
      \midrule
      T5-large & First Token \cite{liu2022separate} & 19.45 & 67.74 & 6.95  & 5.18 \\
      T5-large & FMHA  & 7.15  & 90.96 & 8.09  & 6.89 \\
      T5-base & FMHA  & 8.06  & 90.10 & 7.88  & 6.63 \\
      \bottomrule
      \end{tabular}}
    \label{tab:abla-text}%
  \end{table}%

\begin{table}[t]
    \centering
    \caption{The impact of different parameter sizes on the performance of AudioSpa.}
    \scalebox{1.0}{\begin{tabular}{ccccc}
      \toprule
      Hyperparameters  & MAE ↓ & ACC (\%) & SDR (dB) & SISDR (dB) \\
      \midrule
      $B=30, D=64$ & 6.57  & 91.08 & 8.06  & 6.86 \\
      $B=20, D=64$ & 7.15  & 90.96 & 8.09  & 6.89 \\
      $B=20, D=32$ & 7.72  & 89.99 & 7.45  & 6.11 \\
      \bottomrule
      \end{tabular}}%
    \label{tab:abla-param}%
  \end{table}%

\begin{figure}[t]
    \centering
    \includegraphics[width=0.48\textwidth]{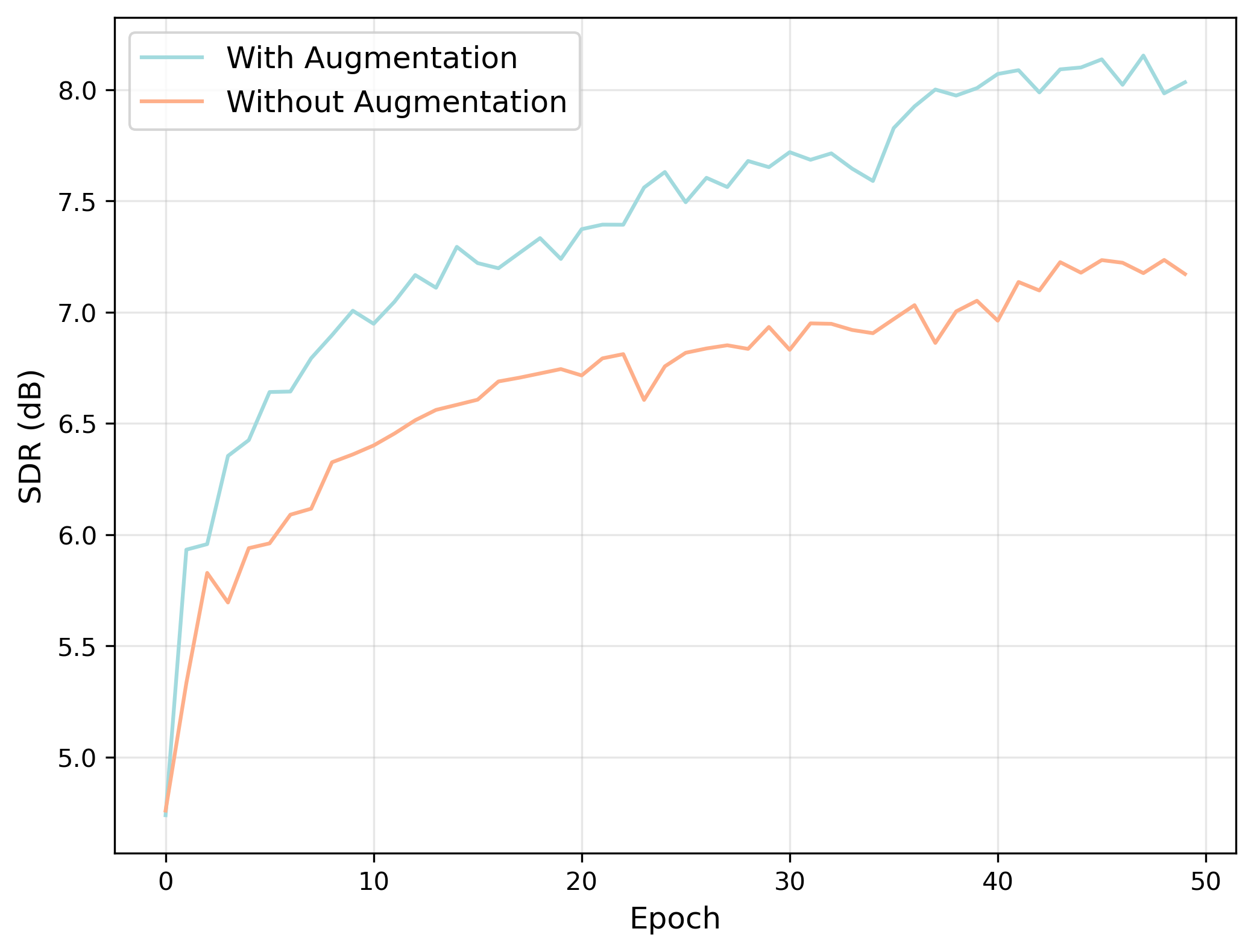}
    \caption{Ablation study of the data augmentation on the validation dataset.}
    \label{fig:aug}
\end{figure}

\begin{table}[t]
  \centering
  \caption{The impact of data augmentation on the performance of AudioSpa.}
    \begin{tabular}{ccccc}
    \toprule
    Augmentation & MAE ↓  & ACC (\%) & SDR (dB) & SISDR (dB) \\
    \midrule
    $\times$ & 7.77  & 87.82 & 7.31  & 5.77 \\
    \checkmark & 6.57  & 91.08 & 8.06  & 6.86 \\
    \bottomrule
    \end{tabular}%
  \label{tab:abla-aug}%
\end{table}%

The first comparison focuses on how textual information is processed. In Table~\ref{tab:abla-text}, the backbone parameters of the model are fixed at $B=20$ and $D=64$. First, we compare token fusion methods. Extracting the first token from the token sequence, as used in \cite{liu2024separate} for text-driven target sound separation, has been proven effective. Our experiments reveal that while this method is useful for spatial audio generation, using FMHA for fusion significantly improves performance, especially in DOA metrics. Compared to directly using the first token, FMHA reduces DOA MAE by 12.3 degrees and increases ACC by 23.22\%. Next, under the condition where FMHA is used, replacing T5-large with the smaller T5-base results in a slight performance decline. Specifically, the DOA MAE increases by 0.91 degrees, and the ACC drops by 0.86\%.

The second comparison focuses on the impact of model parameter size. As shown in Table~\ref{tab:abla-param}, the overall trend indicates that as the model's parameter size decreases, its performance gradually deteriorates. When the residual channel $D$ is fixed at 64 and the number of residual blocks $B$ is reduced from 30 to 20, the performance remains unaffected. Specifically, the DOA MAE increases by only 0.58 degrees, ACC drops by 0.12\%, and SDR and SISDR remain nearly unchanged. However, when the residual channel $D$ is further reduced to 32, the performance degrades noticeably, with SDR and SISDR dropping significantly to 7.45 dB and 6.11 dB, respectively.

The third comparison examines the impact of on-the-fly data augmentation. The backbone parameters of the model are fixed at $B=30$ and $D=64$. As shown in Fig.~\ref{fig:aug}, applying data augmentation techniques to the training data effectively enables the model to learn from a wider range of data distributions, improving its generalization on the validation set. In Table~\ref{tab:abla-aug}, data augmentation results in comprehensive improvements across all evaluation metrics. Specifically, DOA MAE is reduced by 1.2 degrees, classification accuracy increases by 3.26\%, and SDR and SISDR improve by 0.75 dB and 1.09 dB, respectively. These results underscore the importance of expanding the training dataset to enhance model performance.

\subsection{Case Study}
\begin{figure*}[t]
  \centering
  \subfigure[Monaural clean audio]
  {
  \includegraphics[width=0.3\textwidth]{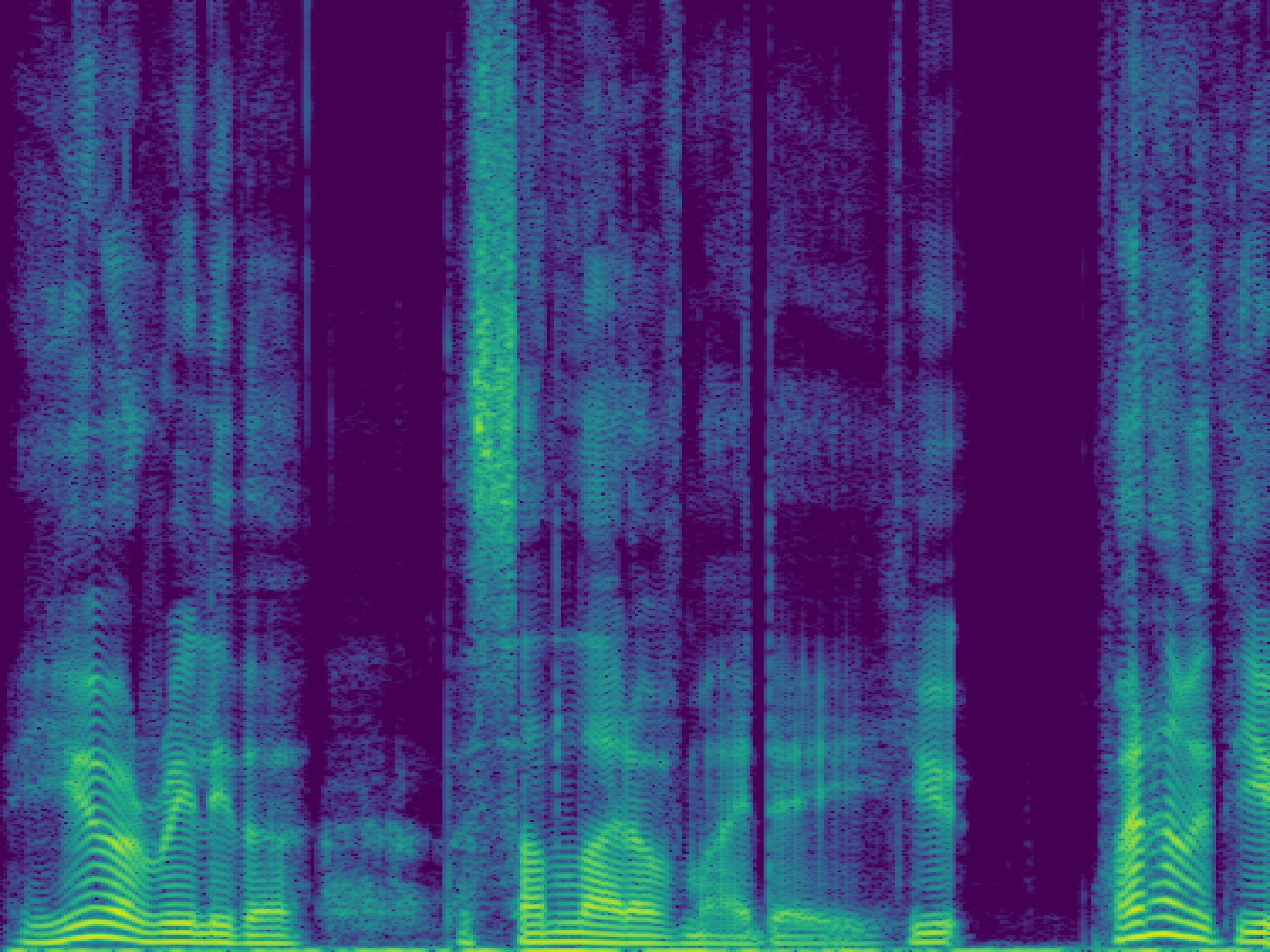}
  \label{subfig:clean_mo}
  }
  \subfigure[Left-received clean audio (AudioSpa)]
  {
  \includegraphics[width=0.3\textwidth]{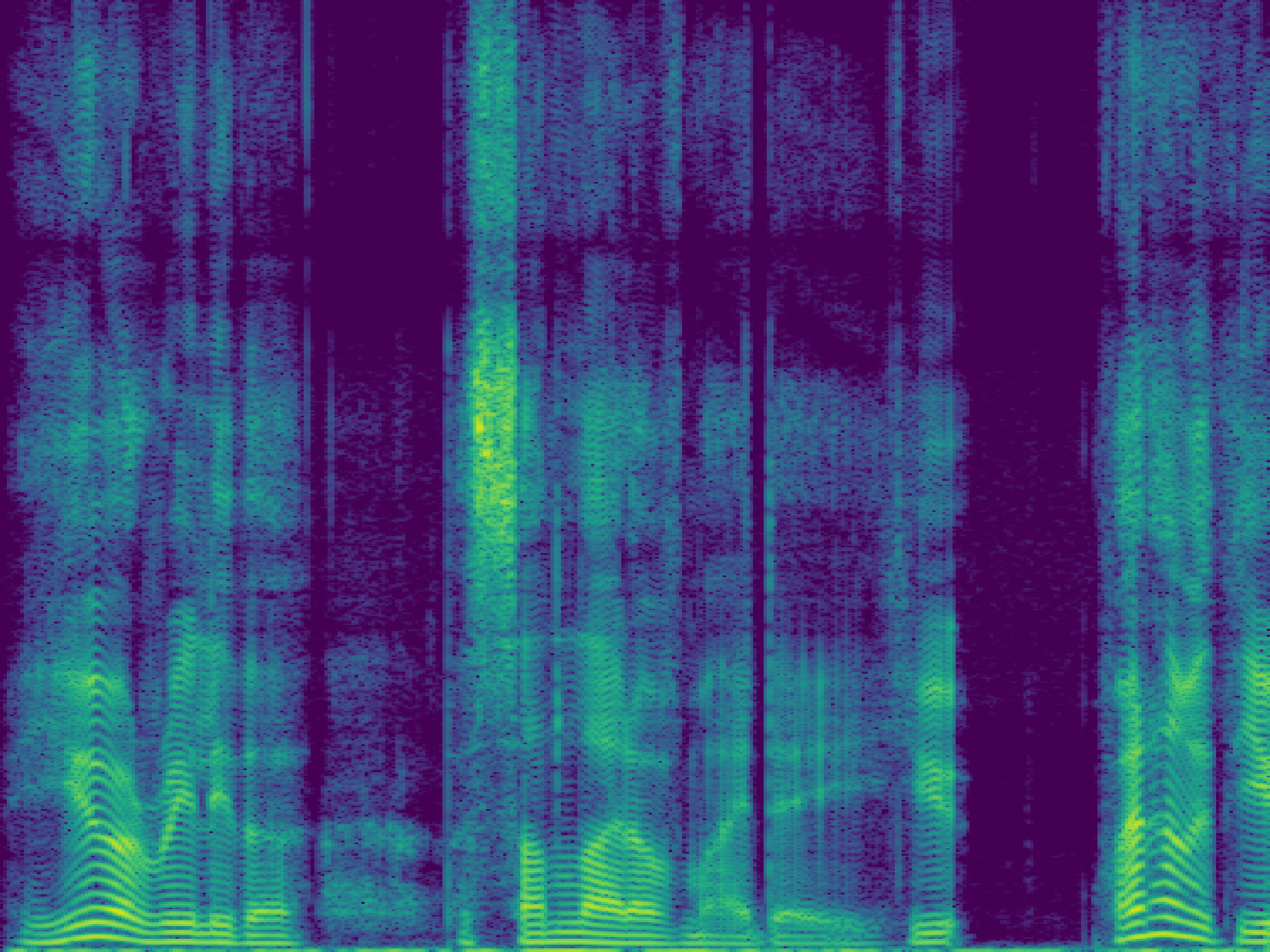}
  \label{subfig:clean_bi_l}
  }
  \subfigure[Right-received clean audio (AudioSpa)]
  {
  \includegraphics[width=0.3\textwidth]{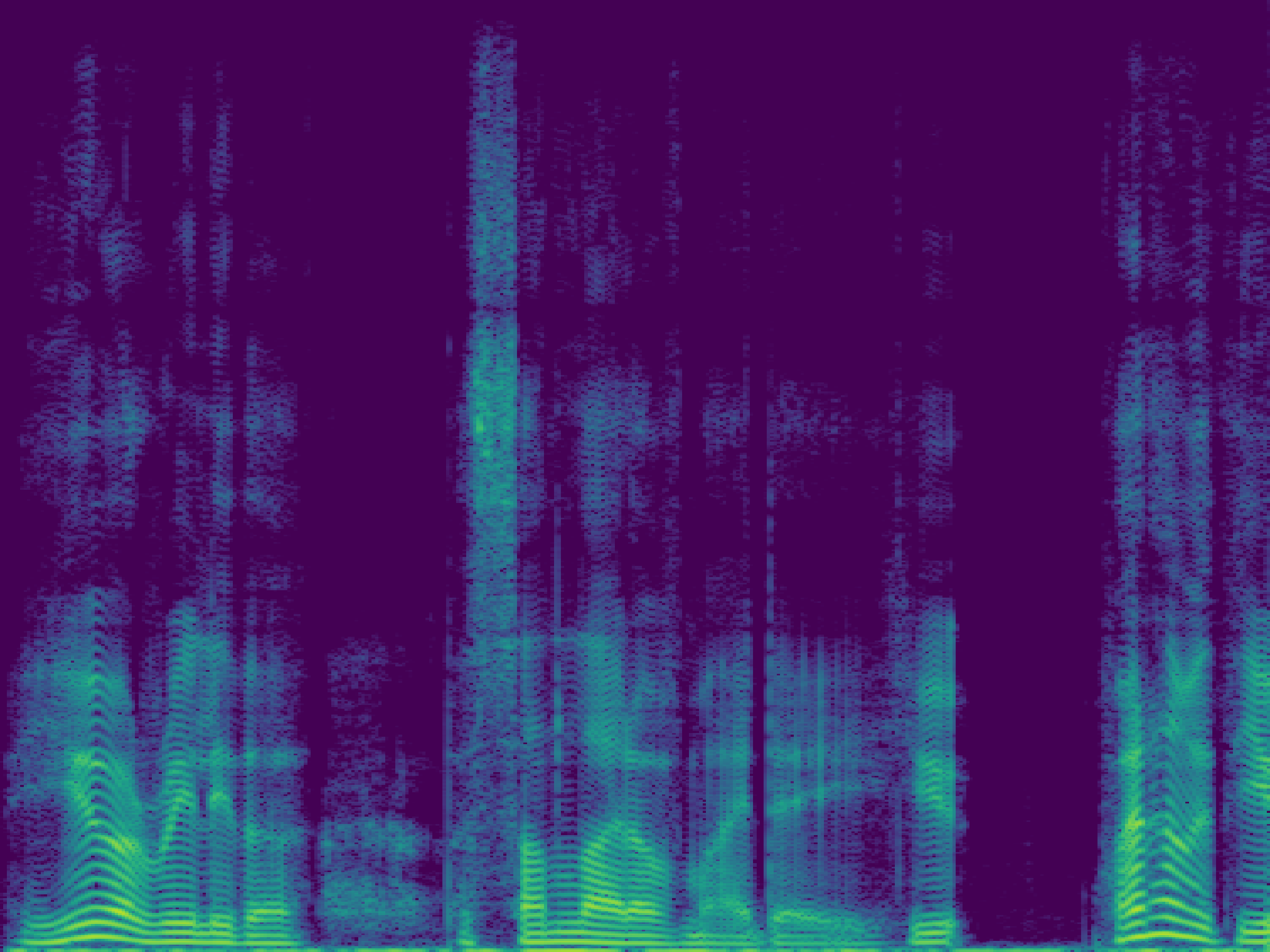}
  \label{subfig:clean_bi_r}
  }
  \subfigure[Monaural noise]
  {
  \includegraphics[width=0.3\textwidth]{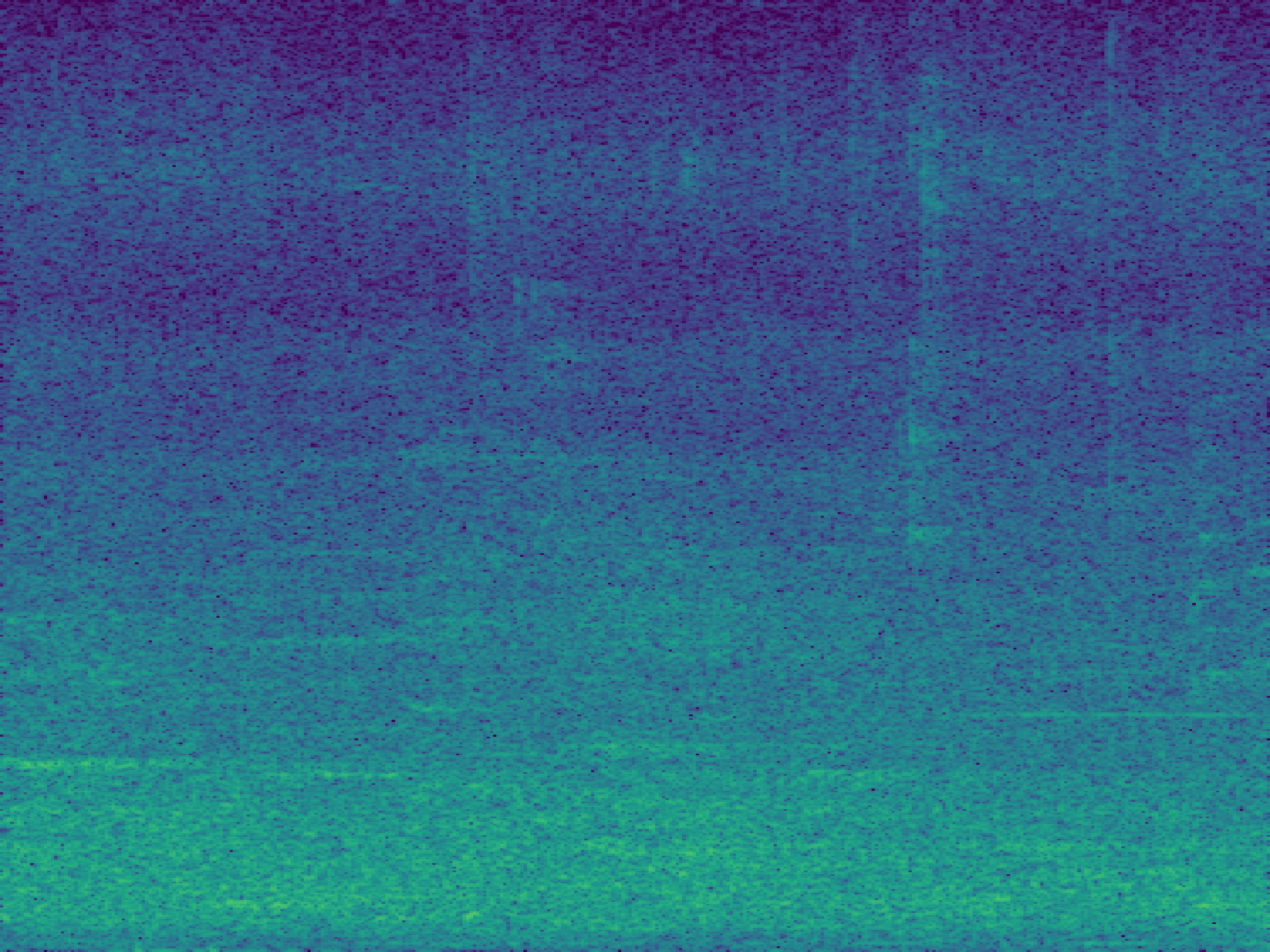}
  \label{subfig:noise_mo}
  }
  \subfigure[Left-received noisy audio (Baseline)]
  {
  \includegraphics[width=0.3\textwidth]{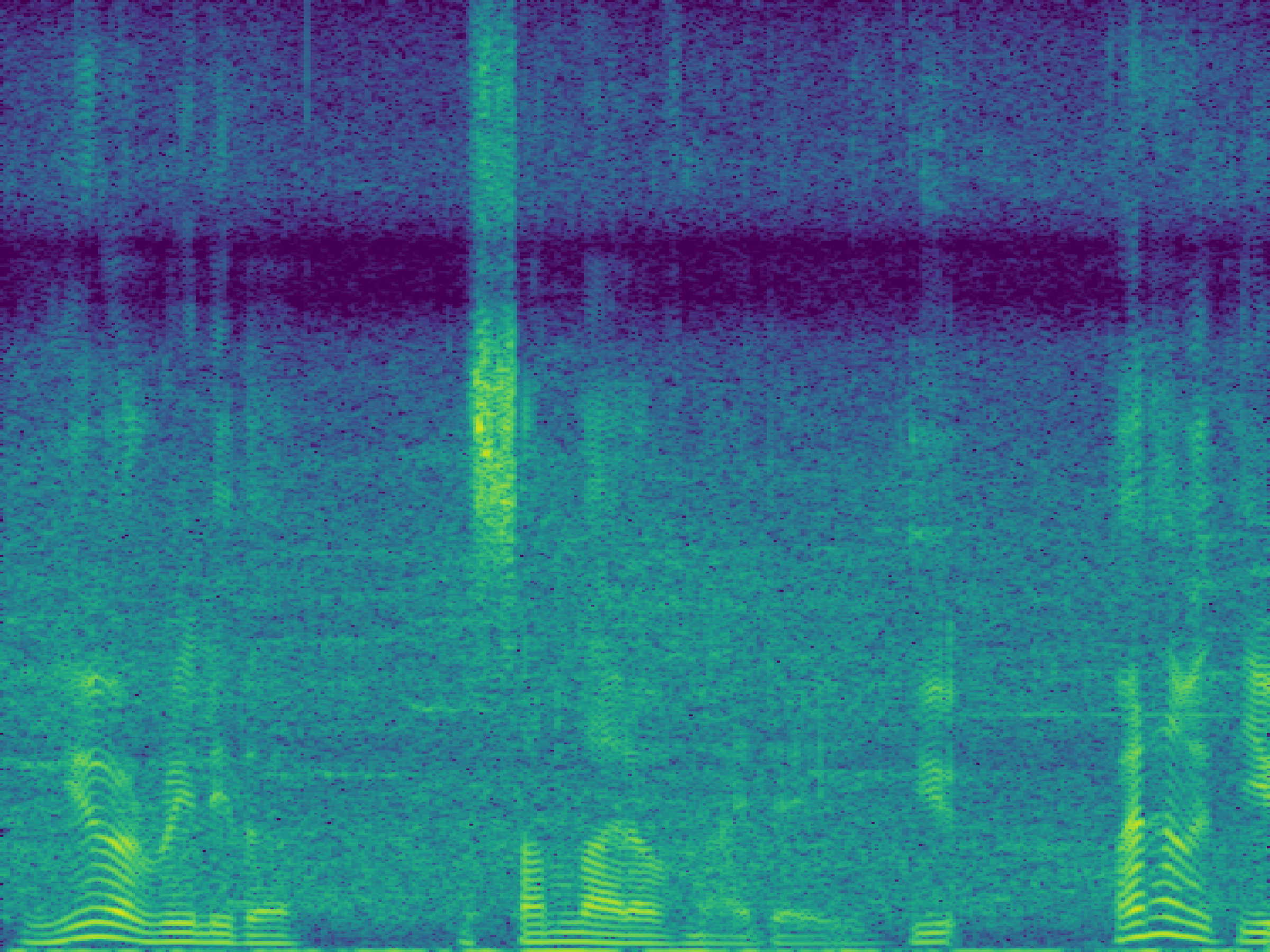}
  \label{subfig:noisy_simu_l}
  }
  \subfigure[Right-received noisy audio (Baseline)]
  {
  \includegraphics[width=0.3\textwidth]{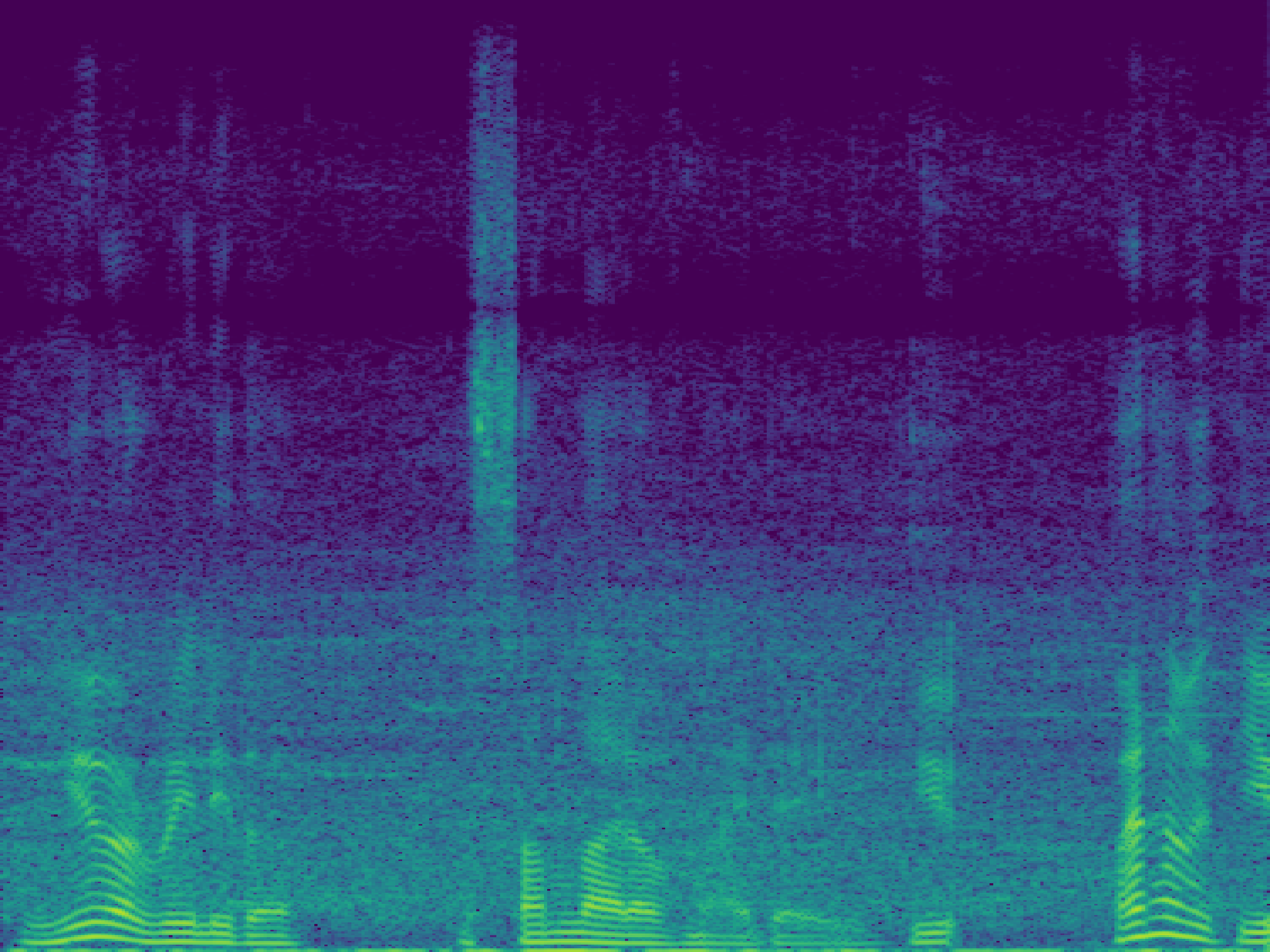}
  \label{subfig:noisy_simu_r}
  }
  \subfigure[Monaural noisy audio]
  {
  \includegraphics[width=0.3\textwidth]{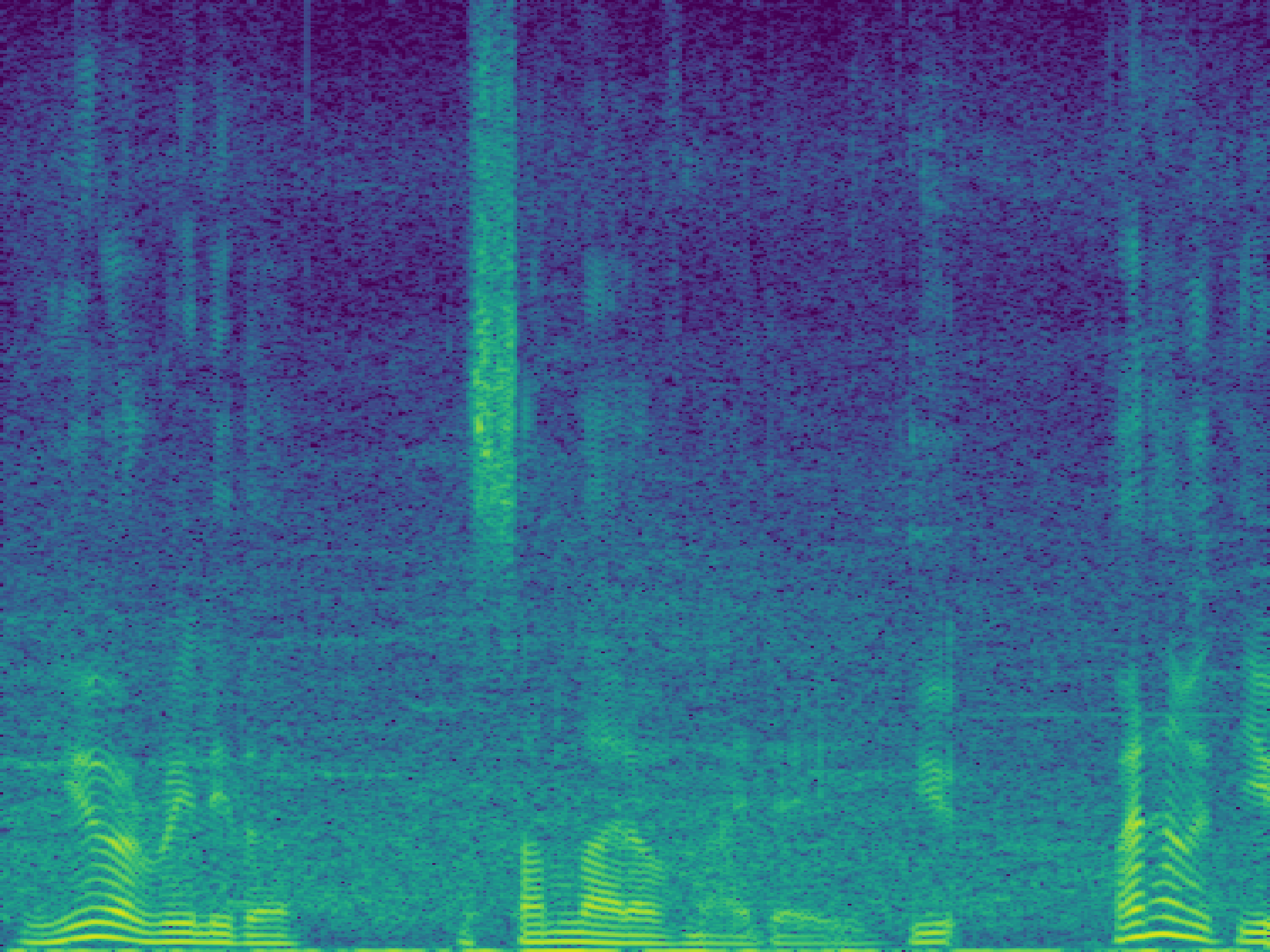}
  \label{subfig:noisy_mo}
  }
  \subfigure[Left-received noisy audio (AudioSpa)]
  {
  \includegraphics[width=0.3\textwidth]{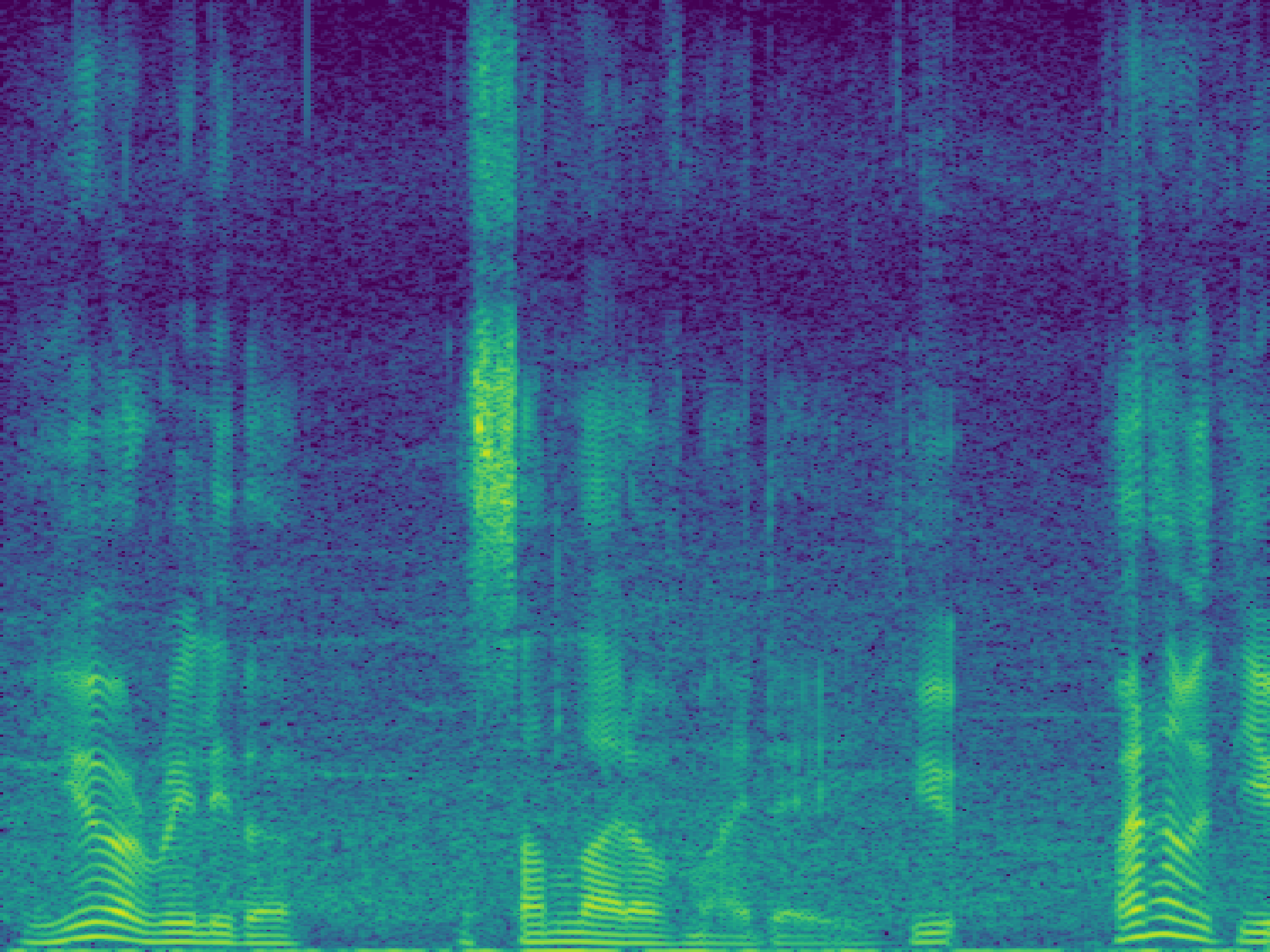}
  \label{subfig:noisy_bi_l}
  }
  \subfigure[Right-received noisy audio (AudioSpa)]
  {
  \includegraphics[width=0.3\textwidth]{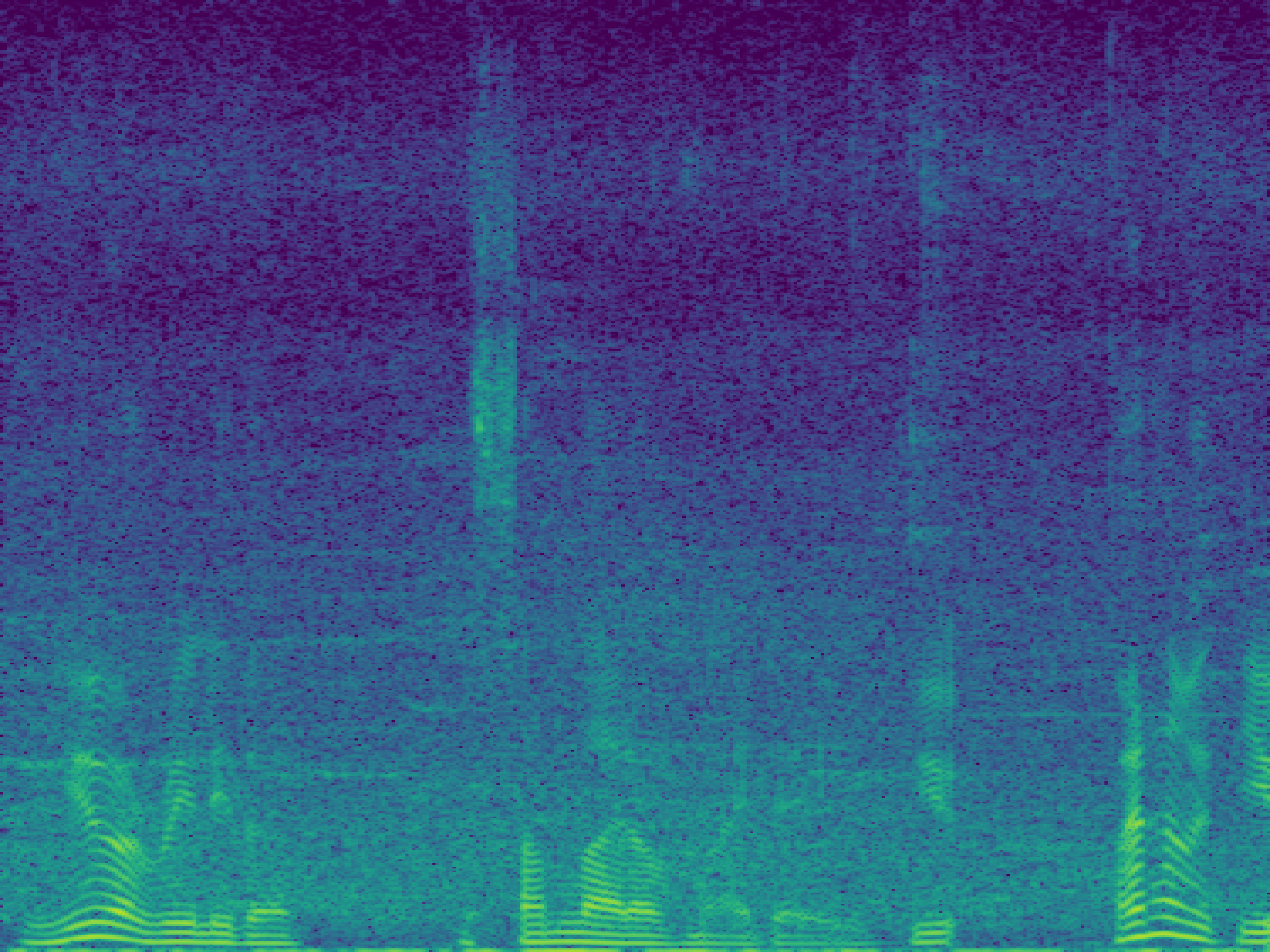}
  \label{subfig:noisy_bi_r}
  }
  \caption{A case study of spatializing specific sound events. Each spectrogram's x-axis represents time, the y-axis represents frequency. Baseline is DSP, SNR is 10dB, and the audio duration is 4 seconds. In the binaural audio, the target sound is positioned directly to the left.}
  \label{fig:one_case}
\end{figure*}

Fig. \ref{fig:one_case} presents a case study illustrating the differences between our AudioSpa and the baseline approach. (a) shows a clean audio signal with a single sound event, while (b) and (c) represent the signals received by the left and right ears, respectively. As the sound is positioned to the left, the left ear receives a noticeably louder signal than the right ear. (d) depicts outdoor urban environmental noise, and (g) is the mixture of (a) and (d). Applying the baseline method to (g) yields (e) and (f), where the right ear signal (f) is significantly quieter, indicating that even the environmental noise is incorrectly localized to the left. This is unreasonable, as such noise should be omnidirectional. In contrast, using AudioSpa on (g) produces (h) and (i). Comparing (g), (h), and (i), it is evident that the environmental noise remains largely unchanged. Furthermore, comparing (b) with (h) and (c) with (i) confirms that the target sound is accurately localized to the left. Thus, our proposed AudioSpa effectively places only the target sound event in the desired direction.

\section{Discussions}\label{sec:dis}

\subsection{Limitation}

\begin{table}[t]
    \centering
    \caption{Results on the two-source data, where the environment is clean.}
    \scalebox{1.0}{\begin{tabular}{ccccc}
      \toprule
            & MAE ↓ & ACC (\%) & SDR (dB) & SISDR (dB) \\
      \midrule
      Ground-truth & 6.11  & 86.90 & $\infty$   & $\infty$ \\
      AudioSpa & 52.40 & 38.30 & 9.39  & 7.64 \\
      \bottomrule
      \end{tabular}}
    \label{tab:2-clean}%
  \end{table}%

In form, the current AudioSpa is capable of accepting monaural audio with two sound sources as input and outputting binaural audio with the two sources spatialized to different directions. We conducted an experiment to test this functionality. We randomly sampled two audio segments from the training set, selected two different HRIRs, and mixed them together to create a training mixture. The environment was clean. All other settings were consistent with those in previous experiments.

As shown in Table~\ref{tab:2-clean}, we first examine the ground-truth audio. The binaural localization model shows some inaccuracy, with a DOA MAE of 6.11 degrees, but it still provides a reasonable reference. For the audio generated by AudioSpa, a weird phenomenon occurs: while the SDR appears fairly good at 9.39 dB, the DOA metrics are quite poor with a MAE of 52.40 degrees. This inconsistency suggests that the current model or loss function may still have room for improvement in future work.

\subsection{Future Work}
In addition to the multi-source sound event issue, there are other challenges in spatial audio generation that remain to be addressed in the future.

(i) Text-to-spatial audio generation without reference audio: The main challenge here is the need to ensure both the quality and accuracy of the generated sound events, as well as ensuring that the spatial perception is accurate. Specifically, current state-of-the-art monaural audio generation models, whether for speech or sound event generation, primarily use Mel spectrograms as features. If other features are used, the model's performance generally degrades. However, Mel spectrograms clearly do not include phase, which is crucial spatial information. In the current work, due to the use of reference audio, the model utilizes time-domain waveforms, which inherently include phase. Future work will likely address this issue through specific design.
% Ideally, a model would be capable of handling both scenarios—one with reference audio and one without. When reference audio is available, the quality of the generated audio would improve, as the model could focus solely on the spatial aspects.

(ii) Visual spatial audio generation: This task would still involve generating spatial audio without reference audio, but the model would need to consider spatial and event references not only from the text but also from other modalities like images or videos. This would offer a more immersive audiovisual experience in future applications.

\section{Conclusions}    \label{sec:con}
In this paper, we propose AudioSpa to address the challenge of text-controlled binaural spatial audio synthesis with monaural reference audio. By integrating a pre-trained large language model and employing fusion multi-head attention to combine token sequences into a single vector, we further utilize this vector for feature modulation within the residual blocks of an acoustic model, enabling text-controlled binaural spatial audio synthesis. During the training phase, we adopted an on-the-fly data augmentation strategy to enrich the data distribution, which significantly enhances the model's generalization capability. We also introduced a binaural localization model to assess the spatial accuracy of the generated audio. Experimental results demonstrate that our proposed algorithm achieves impressive performance in spatial audio synthesis for various single-source scenarios. Additionally, ablation studies validate the effectiveness of both the fusion multi-head attention and the data augmentation techniques. Finally, we made some prospects for future work.

\small
\bibliographystyle{IEEEtran}
\bibliography{Reference}

\end{document}